\documentclass[%
 aip,
 amsmath,amssymb,
reprint,%
]{revtex4-1}
\usepackage{graphicx}
\usepackage{dcolumn}
\usepackage{bm}
\usepackage{blindtext}
\usepackage[utf8]{inputenc}
\usepackage[T1]{fontenc}
\usepackage{mathptmx}
\usepackage{wrapfig}
\usepackage{gensymb}
\graphicspath{{/Figures}}
\usepackage{color}
\usepackage{cancel}
\usepackage{hyperref}
\usepackage{amsmath} 
\usepackage{physics}

\begin{document}

\preprint{AIP/123-QED}

\title[]{Spin- and angle-resolved inverse photoemission setup with spin orientation independent from electron incidence angle}

\author{A. F. Campos}
\affiliation{Université Paris-Saclay, CNRS, Laboratoire de Physique des Solides, 91405, Orsay, France.}

\author{P. Duret}
\affiliation{Université Paris-Saclay, CNRS, Laboratoire de Physique des Solides, 91405, Orsay, France.}

\author{S. Cabaret}
\affiliation{Université Paris-Saclay, CNRS, Laboratoire de Physique des Solides, 91405, Orsay, France.}

\author{T. Duden}
\affiliation{Constructive solutions for Science and Technology, 33649 Bielefeld, Germany.}
 
\author{A. Tejeda}
\email{antonio.tejeda@cnrs.fr}
\affiliation{Université Paris-Saclay, CNRS, Laboratoire de Physique des Solides, 91405, Orsay, France.}

\date{\today}

\begin{abstract}

A new spin- and angle-resolved inverse photoemission setup with a low-energy electron source is presented. The spin-polarized electron source, with a compact design, can decouple the spin polarization vector from the electron beam propagation vector, allowing to explore any spin orientation at any wavevector in angle-resolved inverse photoemission. The beam polarization can be tuned to any preferred direction with a shielded electron optical system, preserving the parallel beam condition. We demonstrate the performances of the setup by measurements on Cu(001) and Au(111). We estimate at room temperature the energy resolution of the overall system to be $\sim170$ meV from $k_{B}T_{eff}$ of a Cu(001) Fermi level, allowing a direct comparison to photoemission. The spin-resolved operation of the setup has been demonstrated by measuring the Rashba splitting of the Au(111) Shockley surface state. The effective polarization of the electron beam is $P=30\pm3$ \% and the wavevector resolution is $\Delta k_{F}\lesssim0.06$ \AA$^{-1}$. Measurements on the Au(111) surface state demonstrate how the electron beam polarization direction can be tuned in the three spatial dimensions. The maximum of the spin asymmetry is reached when the electron beam polarization is aligned with the in-plane spin-polarization of the Au(111) surface state.
\end{abstract}

\maketitle

\section{\label{sec:level1}INTRODUCTION}

The lifting of the spin degeneracy of energy bands in low-dimensional systems is a consequence of the relativistic spin orbit interaction (SOI) and the inversion asymmetry at the surface. The unveiling of the spin character in two-dimensional (2D) spintronic systems is of crucial importance for the manipulation of spin textures without the presence of applied voltages. Fully determining the spin-dependent properties is of major importance on Rashba-type materials \cite{lashell1996spin, hoesch2004spin, hochstrasser2002spin, koroteev2004strong, sugawara2006fermi, Miron2011}, topological insulators\cite{fan2016spintronics, yokoyama2014spintronics, Hasan2010, Qi2011}, spinterfaces \cite{sanvito2010molecular, barraud2010unravelling, Cinchetti2017}, among others. A deep understanding of these systems often requires using a spin-polarized electron beam as a probe. Spin-polarized electron sources are essential, for instance, for spin-polarized low-energy electron microscopy (SPLEEM)\cite{Duden98, Rougemaille10, Quesada21} or spin-polarized inverse photoemission spectroscopy (SPIPES), which is the most direct technique to determine unoccupied electronic states with $k$-resolution \cite{dose1983ultraviolet, smith1988inverse, Himpsel90, Ortega1992, Skibowski94, Themlin1997} and spin-resolution \cite{Donath1989, budke2007combined, cantoni2004high}.

In inverse photoemission spectroscopy (IPES), the electrons emitted from the source couple to unoccupied states above the Fermi level ($E_{F}$) of the sample. Direct transitions to lower-lying unoccupied levels conserve the energy and the parallel-to-the-surface wavevector across the surface, i.e. outside ($\vb{K_{||}}$) and inside ($\vb{k_{||}}$) the solid. If both the direction and the kinetic energy of the incoming electron beam are well defined, the detection of the emitted photons, produced when the optical transition takes place, allows to determine the band structure $E(\vb{k_{||}})$ above the Fermi level. A complete determination of the spin character of an arbitrary electronic state at a particular wavevector needs that the spin polarization vector $\vb{P}$ of the electron beam is fully independent from the electron beam incidence angle $\theta$ with respect to the surface normal. If it is not the case and $\vb{k}=\vb{k}(\theta)$ and $\vb{P}=\vb{P}(\theta)$, the variation of the wavevector $\vb{k}$, necessary to achieve $k$-resolution in IPES, modifies the projection of $\vb{P}$ onto the surface and therefore it is not possible to study an electronic state at an arbitrary wavevector with an arbitrary electron beam polarization. It is thus compulsory to decouple $\vb{P}$ from $\vb{k}(\theta)$ to determine any unoccupied state with both, $k$- and spin-resolution.

\begin{figure}[ht]
	\includegraphics[width=0.4\textwidth]{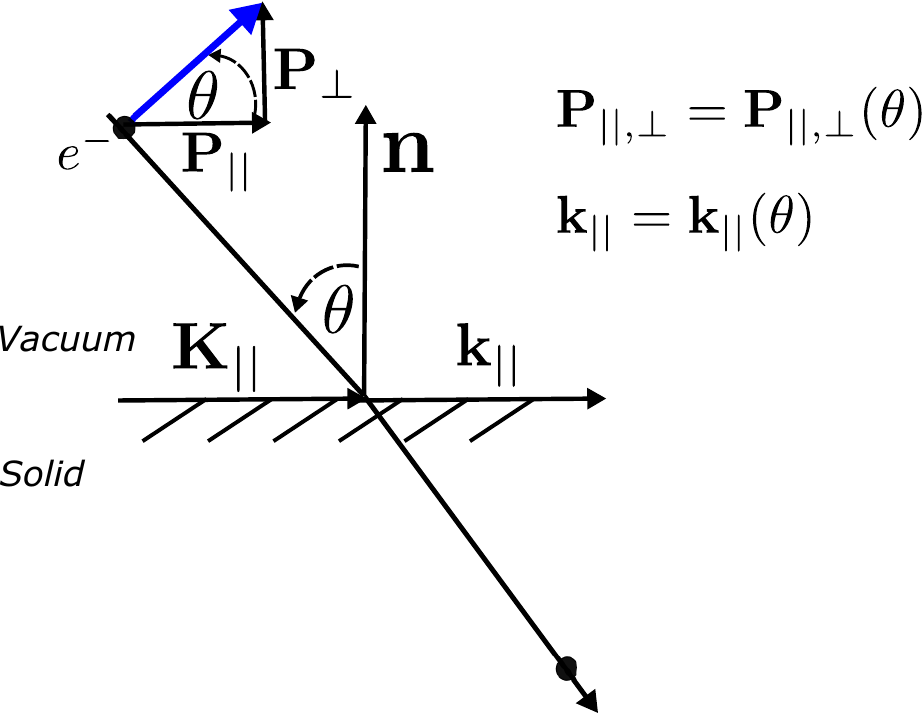} 
	\caption{\label{fig:Fig1} Dependency of $\vb{k_{||}}$ and $\vb{P}$ (blue arrow) on the angle of incidence $\theta$ of the electron beam over the surface. The electron optical system is able to decouple $\vb{P}$ from $\vb{k}$, allowing to maintain the same magnitude of either $\vb{P_{||}}$ or $\vb{P_{\perp}}$ in a $k$-resolved SPIPES measurement.}
\end{figure}

Electron guns operating at low energies are a necessity for spin-polarized inverse photoemission (SPIPES). Current SPIPES-dedicated sources either focus only on the surface component of the polarization or they tune the spin polarization perpendicular to the surface $\vb{P}_{\bot}$ by rotating the sample around an axis at the surface. These setups allow to study a whole variety of systems \cite{stolwijk2014rotatable, wissing2014ambiguity,stolwijk2015spin, datzer2017unraveling,eickholt2018spin, berti2014direct, Ciccacci1997, SJagadeesh2017}. In order to go beyond these state-of-the-art setups, a GaAs photocathode-based electron gun where $\vb{P}$ is independent of the incoming beam angle, has been developed. In this review, we describe the design of a fully three-dimensional (3D) spin source \cite{duden1995compact}, the spin of which can be rotated at will by the use of electron optical elements. The source generates the low-energy electrons necessary for spin-polarized inverse photoemission, with the goal to investigate the unoccupied $E(k)$ dispersion of spin-polarized systems without any restriction on the spin character of the electronic states.

This article is organized as follows. Section \ref{sec:level2} is committed to the experimental generalities of the SPIPES source setup and to the description of the electron optical system that enables the 3D tuning of the electron beam polarization. The energy resolution obtained from IPES measurements on Cu(001) is also discussed. In Section \ref{sec:level3}, the effective polarization and the momentum resolution of the SPIPES spectrometer are quantified from the dispersion of the Au(111) Shockley surface state (SS).  The Rashba splitting of the Au(111) SS allows to demonstrate the rotation of the electron beam spin polarization into longitudinal and transverse orientations.

\section{\label{sec:level2}EXPERIMENTAL SETUP} 

\subsection{\label{sec:level2_0}Ultra-high vacuum setup}
Our spin-polarized electron source is included in an ultra-high vacuum chamber with a $\mu$-metal magnetic shielding and base pressure of $2 \times 10^{-10}$ mbar thanks to a turbo, ion and non-evaporable getter (NEG) pumps. The electron source allows to perform inverse photoemission in isochromat mode. In this operation mode the photons have a constant energy $\hbar\omega$ and the kinetic energy of the electron beam is varied. The distance of the source nose to the sample is about 2 cm. We use Geiger-M\"{u}ller (GM) detectors \cite{dose1977vuv} with CaF$_{2}$ window \cite{thiede2015optimizing} (low-pass filter) and Ar-C$_{3}$H$_{6}$O [96:4] gas blend (high-pass filter). This bandpass characteristic renders a detection energy of $\hbar\omega$ = 9.90 $\pm$ 0.15 eV in each counter.  The GM detectors are installed on linear motion drives to minimize their distance to the sample and to increase the counting rate. A characteristic counting rate of 20 Hz, due to the light of a Bayard-Alpert gauge, was taken as reference of gas stabilization and a dark count rate of about 0.1 Hz was observed in all experiments. The center of the photon detectors are $75\degree$ (GM1) and $35\degree$ (GM2) with respect to the electron gun nose. The gas mixture and the voltage are chosen so that the proportional multiplication mechanism \cite{ thiede2018acetone} is always dominated by the Geiger-M\"uller plateau. The sample manipulator with five degrees of freedom allows to regulate sample temperature between room temperature and 10 K. The analysis chamber is connected to an independent preparation chamber. The preparation chamber allows sputtering and annealing samples, cooling to 77 K, gas exposure, evaporation and sample cleaving. It is also equipped with a quartz microbalance for calibrating evaporations and a LEED/Auger for sample characterization. A secondary independent preparation chamber is currently dedicated to molecular evaporation.

\subsection{\label{sec:level2A}Spin-polarized electron generation}

In the electron source, spin-polarized electrons are photoemitted from negative electron affinity (NEA) GaAs by exciting transitions with circularly-polarized infrared (IR) radiation \cite{scheer1965gaas,pierce1975negative, pierce1976photoemission, pierce1980gaas}. The angular momentum conservation of either right- or left-helicity is used to select the orientation of the spin polarization of the electron beam. For the excitation of the photoelectrons, we use a diode laser with a wavelength of 830 nm and 30 mW of CW power. The laser intensity on the photocathode as well as the helicity of the laser light can be controlled by two liquid crystal (LC) retarders in alternating sequence with two Glan-Thompson polarizers. The first LC retarder is located between two crossed Glan-Thompson polarizers and allows the adjustment of the transmission by its variable birefringence. The second LC retarder receives linearly polarized light from the second Glan-Thompson polarizer. By invoking an adjustable phase shift, which is either set to quarter wave or three-quarter wave delay, a positive or negative helicity is imposed on the outgoing laser beam.

The photocathode is a GaAs(100) crystal with a zinc dopant concentration of $2.5\times10^{19}$ cm$^{-3}$ and front-side illumination. The surface is NEA-activated by annealing to 870 K followed by alternated evaporation of Cs and exposure to O$_{2}$ in a dedicated preparation chamber. After the NEA condition is reached, the cathode is transferred to the operating position in front of the electron extractor where it can be reactivated by a low-flux Cs dispenser which is located in the vicinity. The photocathode lifetime under uninterrupted operation is better than eight hours and it can be extended up to 24 hours with continuous low-flux Cs evaporation.  The quantum efficiency (QE) of the photocathode was estimated as $\sim0.005$ for polarized emission. This value is in agreement with the yield of non-strained GaAs \cite{pierce1975negative, guo2013quantum, ciccacci1991comparative} and high-QE electron sources \cite{sheffield1995high}. 

In SPIPES, it is a common practice to attenuate the laser intensity to minimize the Boersch effect \cite{boersch1954experimentelle} and to preserve the operational lifetime of the photocathode, therefore, current densities below $\sim$ 0.8 $\mu$A$\cdot$mm$^{-2}$ were used by adjusting the power of the laser. All the presented  spectra are normalized to the incident electron beam current at the sample unless otherwise stated. The emitted current from the photocathode and the transmitted current to the target can be monitored in dedicated electrometers with an electrical resolution of about 50 pA. The transmission of the electron source can be optimized up to 70\% for a purely electrostatic deflection and about 55\% for any of the magnetic deflection cases (see Section \ref{sec:level2B}) at the energy regime of (5-20) eV.

\begin{figure}[ht]
	\includegraphics[width=0.5\textwidth]{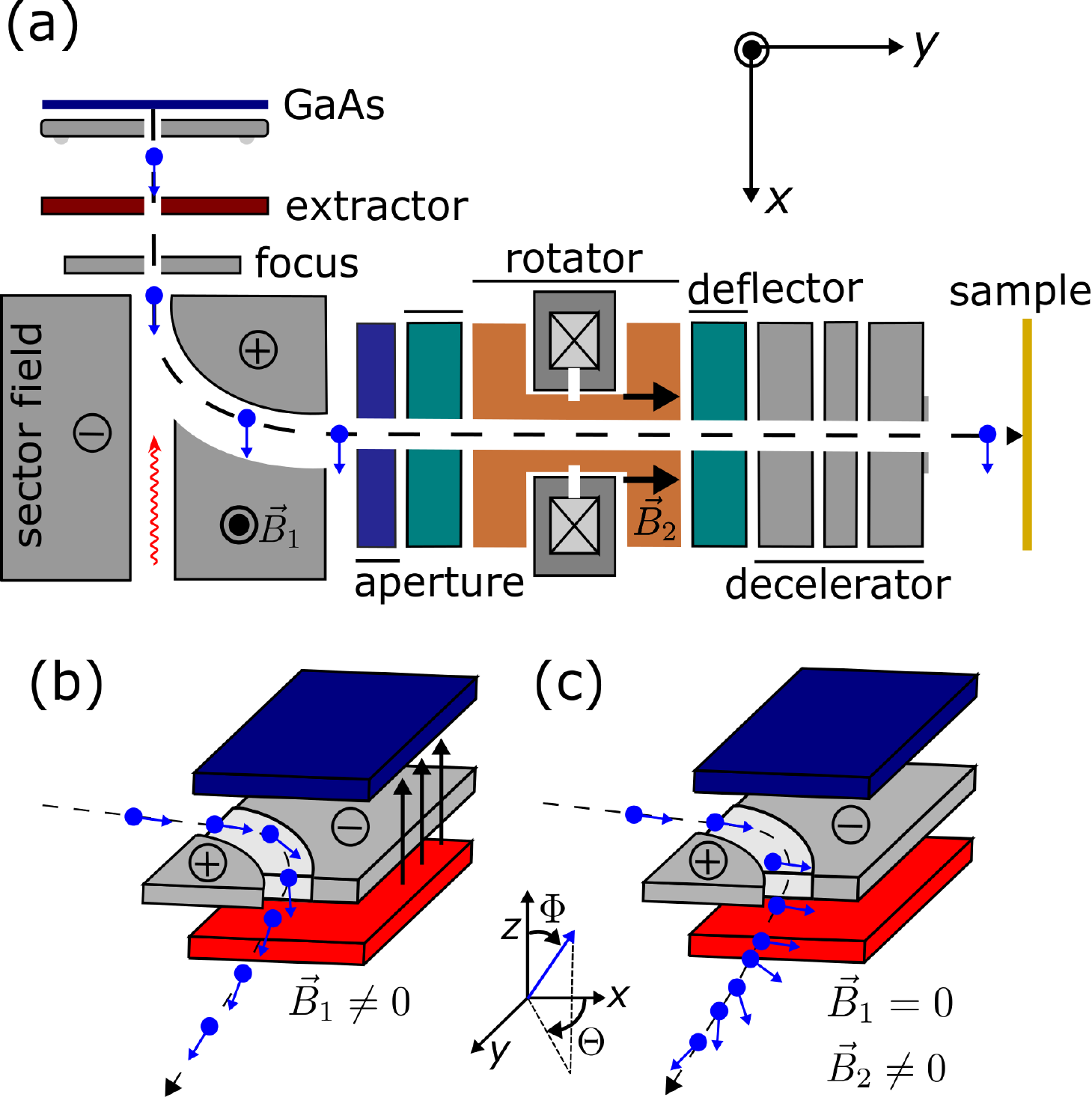} 
	\caption{\label{fig:Fig2} Simplified schematics of the low-energy spin-polarized electron source. (a) Photoemission from NEA-activated GaAs is produced with a circularly polarized IR beam. Electrons are extracted with a voltage of 2 kV in a Ti electrode and subsequently deflected by an electromagnetic sector field. The beam polarization $\vb{P}$ is controlled by the magnetic field applied in the sector field. (b) $\Theta$-rotation: a total magnetic deflection from the polepiece in the sector field can preserve the spin polarization along the propagation axis ($\vb{P}\parallel \vb{v_e}$). (c) $\Phi$-rotation: in a purely electrostatic deflection the relation $\vb{P}\perp \vb{v_e}$ holds after the sector field but a subsequent precession of $\vb{P}$ ($\vb{B_2}\neq \vb{0}$) allows to select any component perpendicularly to the propagation axis.} 
\end{figure}

\subsection{\label{sec:level2B}Electron-optics}

In most spin-polarized electron source designs with front illumination, an electrostatic sector field is following the cathode. Thus, the angle between the polarization $\vb{P}$ and the beam propagation vector is limited only to transverse spin polarization. In inverse photoemission, the reciprocal space wavevector $\vb{k}$ is tuned by changing the incident angle $\theta$. The geometric relation between $\vb{P}$ and $\vb{k}$ is shown in Fig. \ref{fig:Fig1}. It is obvious that, if $\theta$ is changed in order to reach a certain point in reciprocal space, also the projection of $\vb{P}$ onto $\vb{n}$ is varied. With this coupling between parameters, there are reciprocal space regions which are difficult to explore e.g., states close to $\Gamma$ (normal incidence) with an out-of-plane spin orientation.

In the following, we present an electron source that overcomes these limitations. The principle has been previously described \cite{duden1995compact}. Here we describe its design in detail. In Fig. \ref{fig:Fig2}(a) it is shown that the photoelectrons are initially emitted with $\vb{P}$ normal to the photocathode surface, with the helicity vector of the light source defining the quantization axis. The extractor accelerates the electrons towards a compact combined $90\degree$ electromagnetic sector field. By the superposition of electrostatic and magnetic deflecting fields in the sector, the angle between $\vb{P}$ and $\vb{k}$ of the transmitted electron beam can be tuned without changing the direction of the outgoing beam\cite{Meister1962}. Before entering the sector field, the electron velocity  $\vb{v_e}$ is perpendicular to a tunable magnetic field $\vb{B_1}$ as depicted in Fig. \ref{fig:Fig2}(b). In the case of a total magnetic deflection, $\vb{P}$ becomes parallel to the momentum vector and the electron beam polarization is longitudinal. Conversely, in a purely electrostatic deflection, the initial orientation of $\vb{P}$ is preserved in the laboratory frame, resulting in a transverse polarization, as depicted in Fig. \ref{fig:Fig2}(c).  In this way, after the $90\degree$ circular sector, the angle $\Theta$ between the electron beam polarization and the propagation vector is controlled in between the limiting cases (i) $\vb{P}\parallel \vb{v_e}$ and (ii) $\vb{P}\perp \vb{v_e}$. 

The electromagnetic sector field is followed by an aperture that only selects the electrons close to the beam center. Thereby, stray electrons are removed from the beam profile which would travel too far from the axis to be properly transferred through the following electron optics. In order to render the beam less sensitive to stray magnetic fields, the transfer energy is set to 1 keV in most parts of the electron optical system, which is, for the same reason, housed inside a $\mu$-metal screen. The selected transfer energy also aims at reducing the effects of space-charge at the electron source, which could have a negative effect on the electron transmission\cite{maniraj2014influence, stoffel1985low}.

\begin{figure}[ht]
	\includegraphics[width=0.5\textwidth]{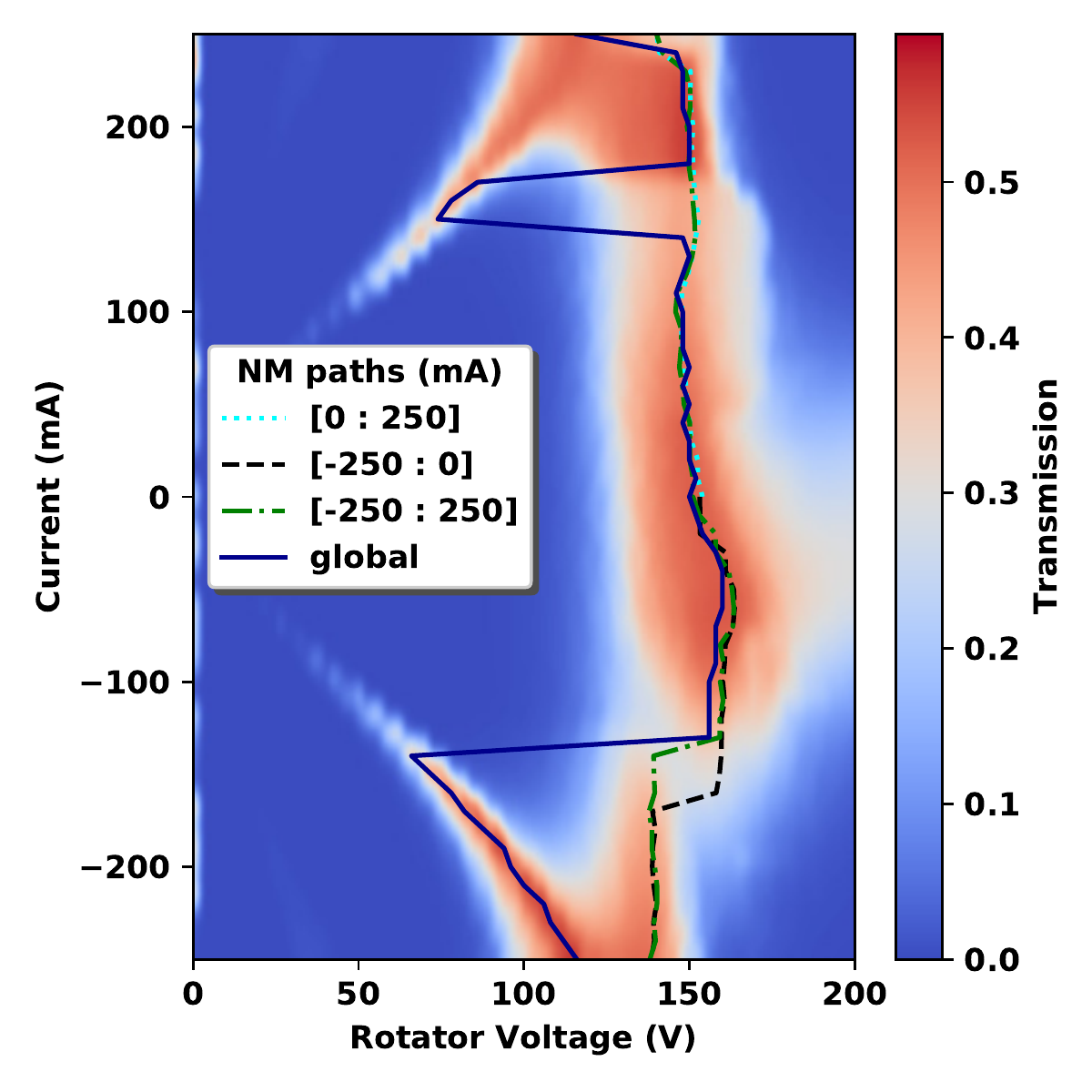} 
	\caption{\label{fig:Fig3} Electron transmission ($E_{kin}=10$ eV) as function of the rotator voltage and the current on the rotator coils. The NM method was applied to the interpolated (bicubic) data by varying the current (mA) along the paths: [0:250], [-250:0] and [-250:250], represented with dotted-cyan, dashed-black, and dot-dashed green lines, respectively. The transmission converges to the global maxima (blue line) in the three paths for the region [130:170]V.}
\end{figure}

The aperture is followed by a rotator lens (Fig. \ref{fig:Fig2}(a), orange) that is embedded between two electrostatic lenses (Fig. \ref{fig:Fig2}(a), green) to preserve the location of the focal planes behind it and to form a focal spot on its center. Thus, the trajectories of the electron beam are all center rays of the rotator lens, greatly unaffected by changes of its magnetic excitation. Transverse polarization components of $\vb{P}$ precess around the axial field $\vb{B_2}$ of the rotator lens \cite{Meister1962} with an angular rotation defined by $\Phi$. The Larmor precession of the electron enables either clockwise or counterclockwise rotations whose magnitude depends on both the current in the rotator solenoid and the rotator lens potential. The rotator potential is set to maximize the transmitted current onto the target alongside a tight focusing of the beam. Both conditions were experimentally explored  by means of the heuristic Nelder-Mead (NM) method \cite{nelder1965simplex}. The NM method provided with optimized voltages on the rotator for maximizing the electron transmission by using the current in the rotator coils as a dependent parameter. Fig. \ref{fig:Fig3} exhibits the electron transmission as a function of the current and the voltage of the rotator for a 10 eV electron beam. The symmetry of the rotator voltage with respect to the initial condition ($\vb{B_2}=\vb{0}$) and the parameters for a maximum target transmission are evidenced. 

After the rotator, the electrons travel through the decelerator stage in which their kinetic energy is reduced to the values desired for the experiment. In the same instance, the decelerator provides the parallel beam condition for the transmitted electrons. The last electrode of the electron gun is held at the same potential as the sample. Thus, the electron beam can travel to the sample in a field-free region which also allows for angle resolved inverse photoemission by sample tilting. In our case, the $\vb{P} (\Theta, \Phi)$ vector of the quasimonoenergetic electron beam is fully decoupled from ${\vb{k}}(\theta)$.

\section{\label{sec:level3}SOURCE PERFORMANCE}

\subsection{\label{sec:level2C}Energy resolution}

\begin{figure}[ht]
	\includegraphics[width=0.5\textwidth]{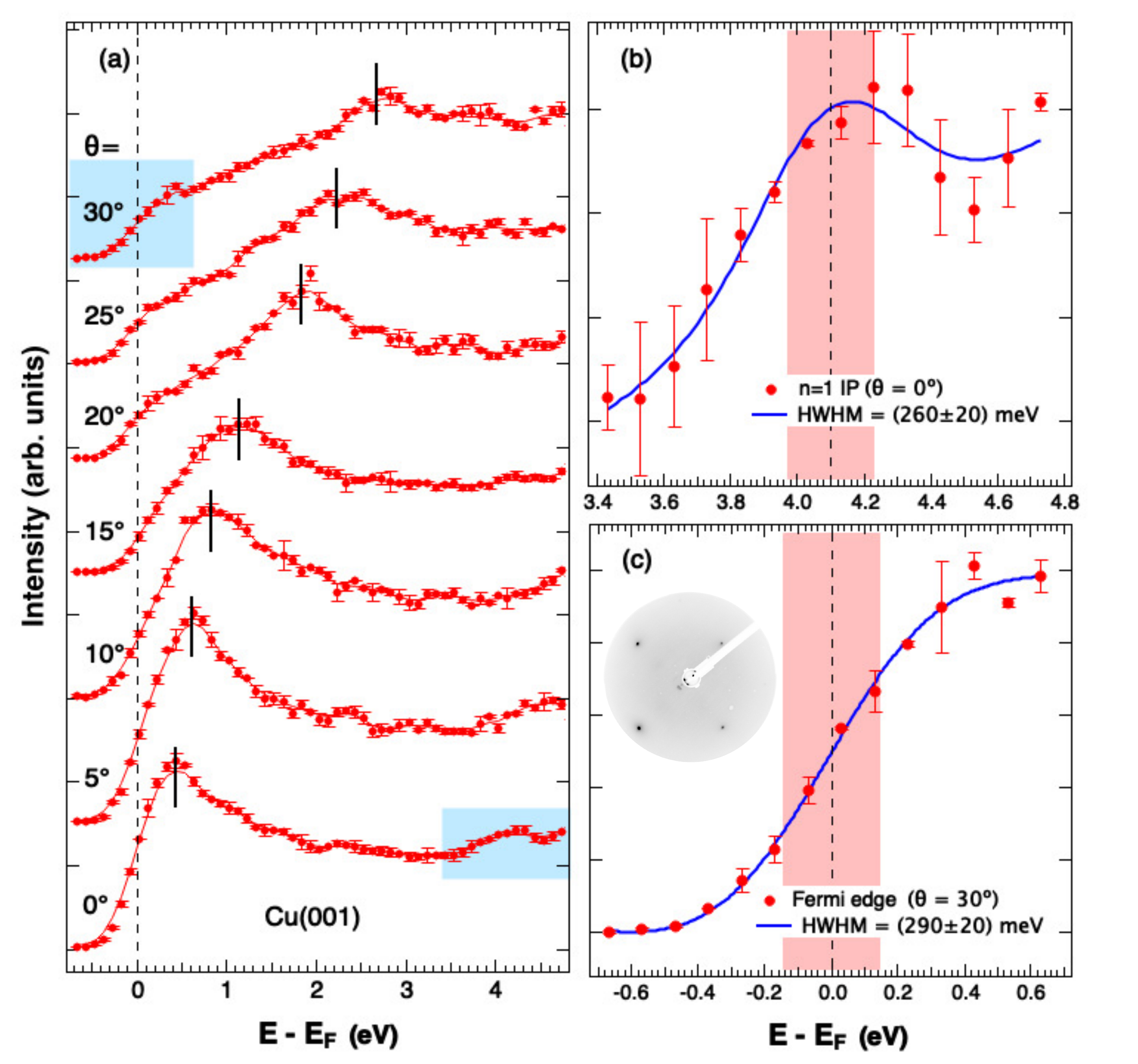} 
	\caption{\label{fig:Fig4}  (a) Room temperature IPES spectra of Cu(001) in the $\Gamma XUL$ plane. The data is depicted by filled-red circles with uncertainty bars. The free-electron-like dispersion of the bulk $sp$-transition is marked with black bars. Notice that the dispersion is unaffected by background subtraction and relative spectral intensities. The blue rectangles delimit the spectra of: (b) the $n=1$ IP state ($\theta = 0\degree$) and (c) the Fermi edge ($\theta = 30\degree$), with their respective spectral fitting (blue lines). Inset: corresponding LEED pattern at 70 eV.}
\end{figure}

Direct and inverse photoemission have estimated energy resolution of experimental setups by looking at different observables. Inverse photoemission has often adopted the Full Width at Half Maximum (FWHM) of the gaussian convoluting Fermi edges \cite{budke2007combined, budke2007inverse} or the FWHM of Image Potential States  \cite{straub1984identification, donath1986photon, budke2007combined}. The FWHM of a gaussian distribution corresponds to $2 \sqrt{2 \ln{2}} \sigma$, with $\sigma$ being the standard deviation. Thus, $\sigma$ is closer to the HWHM of the gaussian distribution (red shaded areas in Fig. \ref{fig:Fig4}) but also to the effective thermal energy $k_{B}T_{eff}$ resulting from the Fermi edge fit to a Fermi-Dirac function, as will be shown below. In the following, we make use of three methods for obtaining observables related to the energy resolution. We propose the use of $k_{B}T_{eff}$ to compare the resolution of both direct and inverse photoemission setups since it can be estimated without any fitting by measuring the width of the Fermi-Dirac function between 10 and 90\% intensity, which corresponds to a width of $4k_{B}T_{eff}$.

One way to compare the performance of our setup to others is achieved by evaluating the spectral width of the image potential (IP) states of Cu, that have a small particle lifetime \cite{royer1988refracting, stiepel2005vacuum, budke2007inverse}. We performed IPES measurements on a freshly prepared Cu(001) surface sputtered with Ar ions (1 keV) and annealed at 770 K. The angle of incidence $\theta$ was varied in the $\Gamma XUL$ plane as determined by LEED. Each data point integrates the raw photon counts during 30 s so that the acquisition time per spectrum is about 30 min. The resulting spectra, normalized by the target current, is presented in Fig. \ref{fig:Fig4}(a) with error bars that correspond to the standard deviation in different sweeps. Binomial smoothing \cite{marchand1983binomial} was performed to include linewidths with reduced statistical noise as a guide to the eye.  Fig. \ref{fig:Fig4}(b) zooms-in the Rydberg-like $n=1$ IP state at normal incidence ($k_{||} = 0$). The state was fitted with a gaussian over a linear background. The Half Width at Half Maximum (HWHM) is $ 260\pm20$ meV, in good agreement with previous findings \cite{straub1984identification, donath1986photon, budke2007combined}.

A second way of estimating the performance of our setup is to analyze the Fermi edge spectral shape. IPES intensity $I(E)$ can be expressed close to $E_{F}$ as the convolution of a Fermi-Dirac distribution for unoccupied states $f_{D}(E,T)$ with a gaussian-like apparatus function \cite{stolwijk2014rotatable,eickholt2018spin} $G(E)$ as:

\begin{equation}
    I(E) = G(E) \star [f_D(E,T)\times B(E)].
    \label{eq:SignalS}
\end{equation}

\noindent where $B(E)=a+bE$ is a weighting factor consisting of an intrinsic constant background due to dark counts on the detector and the almost-constant DOS of a free-electron metal \cite{budke2007inverse}. The total energy resolution of the spectrometer is often considered to be the FWHM of the apparatus function in Eq. \ref{eq:SignalS}. We have measured on the Fermi level of Cu(001) the band crossing at $\theta = 30\degree$ (Fig. \ref{fig:Fig4}(c)). At this angle the bulk $sp$-band is well above the Fermi level, so the width of the apparatus function can be obtained by a fit to the IPES signal close to $E_F$. The HWHM of the apparatus function appears to be $290 \pm 20$ meV, comparable to other setups \cite{budke2007combined, budke2007inverse}.

Finally, a straightforward way to compare the experimental resolutions of different experimental setups, valid also for photoemission, is to evaluate the effective thermal energy $k_{B}T_{eff}$ of the experimental Fermi level that includes all the broadening sources in the setup. In order to obtain a value without any mathematical treatment, instead of fitting to a Fermi function to determine $k_{B}T_{eff}$, we have measured the width of the Fermi edge between 10\% and 90\% of its intensity, obtaining that $4k_{B}T_{eff} \sim 680$ meV and therefore $k_{B}T_{eff}$ is 170 meV. 

\begin{figure}[ht]
	\includegraphics[width=0.5\textwidth]{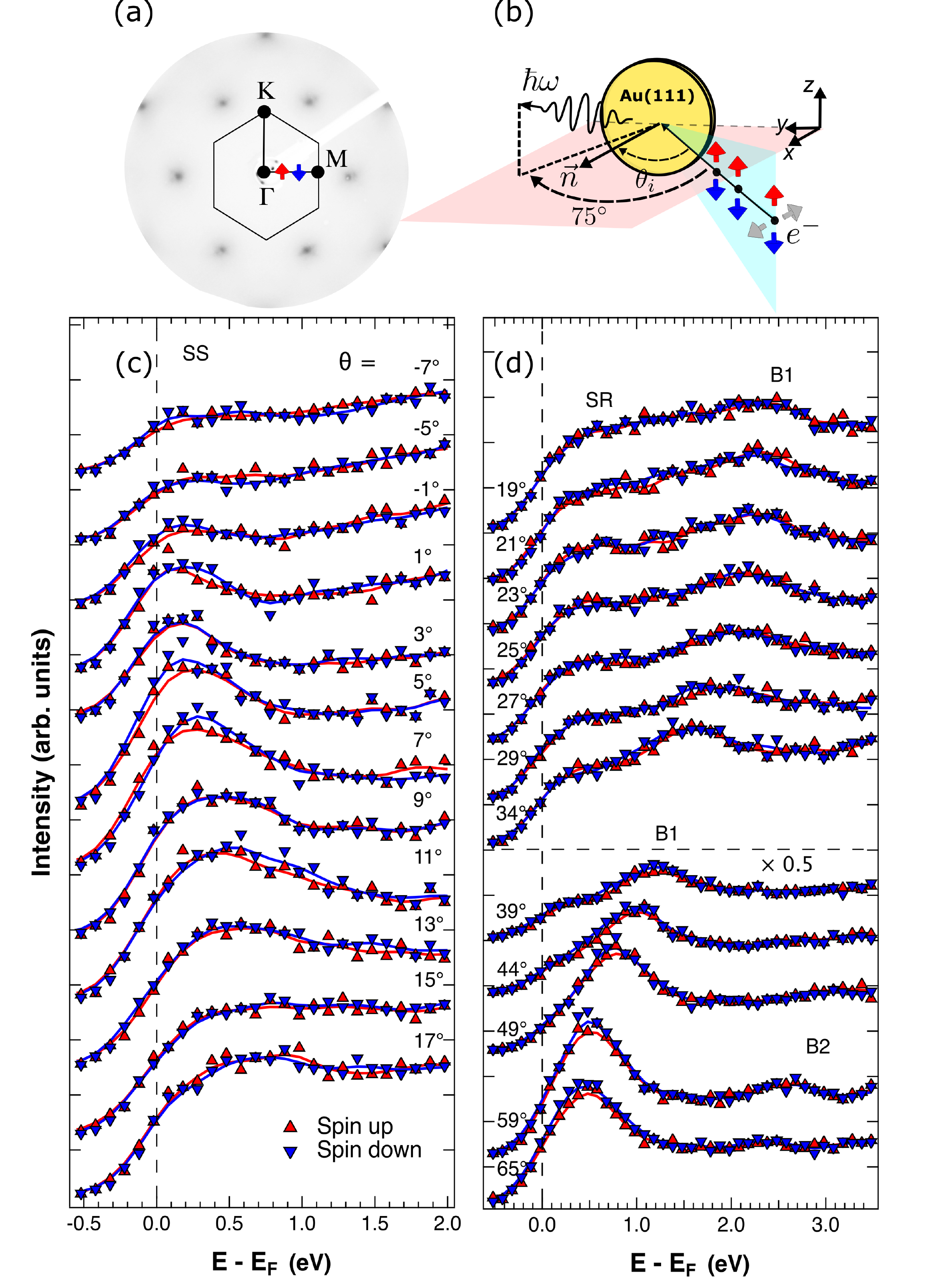} 
	\caption{\label{fig:Fig5} (a) LEED pattern of Au(111) at 150 eV with sixfold symmetry. The up-red (down-blue) arrow along $\overline{\Gamma M}$ depicts the spin up (down) component of $\vb{P}$. (b) Experimental geometry for the Au(111) SPIPES spectra acquired with the GM1 detector. Room temperature SPIPES on Au(111) along $\overline{\Gamma M}$ showing (c) SS, (d) SR and bulk states (B1 and B2). The data have been normalized to the beam current on the sample and the spectral intensities are scaled to 0.5 for $\theta \geq 38\degree$.  Corresponding smoothed linewidths serve as a guide to the eye.}
\end{figure}

\subsection{\label{sec:level3_0}Spin polarization resolution}

We have analyzed the spin resolution of our system by determining the electronic structure of Au(111). At the surface, the lack of inversion symmetry gives rise to a spin-split SS at the L-gap consistent with the Rashba-Bychkov spin-orbit coupling of 2D gases \cite{henk2004spin,hoesch2004spin}. The Au(111) surface was prepared by Ar ion sputtering (1 keV) and annealing at 800 K. Surface quality and orientation were determined by LEED (Fig. \ref{fig:Fig5}(a)). The $\overline{\Gamma M}$ direction was selected for measurements to avoid any influence of the herringbone reconstruction on the SS \cite{reinert2004influence}.

\begin{figure}[ht]
	\includegraphics[width=0.4\textwidth]{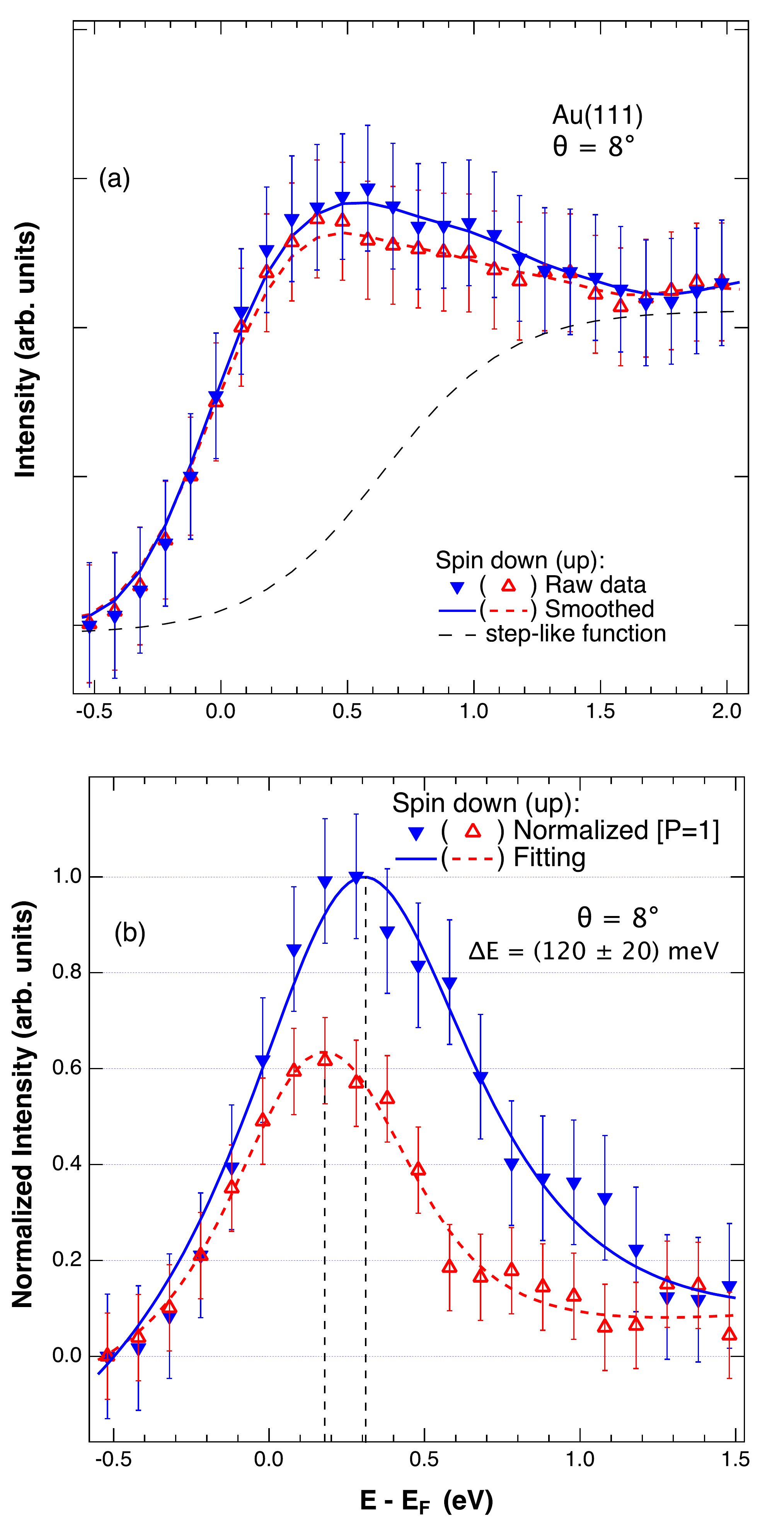} 
	\caption{\label{fig:Fig6} Spin-split SS at $\theta = 8\degree$ with the spin down (up) component depicted by solid (empty) circles. The data have been normalized to the beam current on the sample. (a) Raw data with non-zero spin asymmetry. (b) Normalized data to a totally polarized beam ($P=1$) and pseudo-Voigt fitting (lines). Normalizing to a 100\% spin polarization always has the effect of clarifying the spin asymmetry \cite{stolwijk2014rotatable}. A step-like background has been subtracted from the raw data\cite{eickholt2018spin}. Error bars correspond to the standard deviation at each point. }
\end{figure}

Room temperature SPIPES was performed using the experimental geometry of Fig. \ref{fig:Fig5}(b) with the GM1 counter. The GaAs photocathode was illuminated with IR light of positive and negative helicity to invert the spin polarization of photoemitted electrons with an integration time of 30 s per data point. The raw photon counts were normalized to the sample current as summarized in the spectra of Fig. \ref{fig:Fig5}(c)-(d). Fig. \ref{fig:Fig5}(c) shows the SS dispersion while Fig. \ref{fig:Fig5}(d) shows the surface resonance (SR) and the bulk states labeled as B1 and B2. The spectra are in qualitative agreement with earlier studies \cite{wissing2013rashba,zumbulte2015momentum,woodruff1986empty} though with lower signal/noise ratio. The $E(k)$ dispersion extracted from Fig. \ref{fig:Fig5}(c)-(d) is shown in the Supplementary Material. Notice that background subtraction and relative spectral intensities are not relevant for obtaining $E(k)$ dispersions. For  quantitative comparison, we have therefore measured one of the more pronounced features in the SS at $\theta=8\degree$ with increased statistics (40 Hz counting rate and 60 s per data point). Fig. \ref{fig:Fig6}(a) shows the spectra, after normalization to the target current. The standard deviation is included as error bars that overlap only partially at the binding energy of the SS, indicating a non-zero asymmetry precisely at the SS. The error bars included in Fig.  \ref{fig:Fig6} cannot be compared to the literature \cite{budke2007combined, stolwijk2014rotatable,wissing2014ambiguity,stolwijk2015spin,eickholt2018spin}. More information can be extracted after normalizing the spectra to the effective spin polarization of the electron beam (see Section \ref{sec:level3A}).  The normalized spectra were fitted by pseudo-Voigt functions over linear backgrounds \cite{lashell1996spin} with a step function centered on the binding energy of the state \cite{eickholt2018spin}. The fitting corresponds to lines in Fig. \ref{fig:Fig6}(b). Our results reproduce the spin asymmetry of the SS in previous studies \cite{wissing2013rashba}. Moreover, we find an energy splitting of $\Delta E \sim 120\pm 20$ meV, also in agreement with the experimental splitting (110 meV) and the theoretical estimation ($\sim$150 meV) \cite{lashell1996spin}.

\subsection{\label{sec:level3A}Effective polarization}

In a GaAs photocathode, the splitting of the valence band into heavy- and light-hole bands at $\Gamma$ imposes a theoretical limit to the polarization of the photoemitted beam, which is $P=0.5$ \cite{pelucchi1995spin,pierce1975negative}. However, depolarization mechanisms can considerably diminish this theoretical value \cite{erbudak1978depolarization,spezia2010temperature, allenspach1984spin} so that the experimental polarization is commonly in between 0.2 and 0.35 at room temperature.  In order to estimate the polarization $P$ of the electron source, we have compared our measurements to those of the literature\cite{donath1989spin} with a $P$-calibrated electron source \cite{wissing2013rashba}. In setups with similar energy resolution and similar experimental geometry of the photon detectors, the intensity ratio between spin down and spin up components should be the same. If we focus on the B1 state of Au(111), the highest asymmetry is reached at $56\degree$ \cite{wissing2013rashba}. If the spin up spectrum is normalized to the spin down one, only the intensity of the former must be reproduced to determine the spin polarization of the electron source (see the Supplementary Material). The state was again fitted by a pseudo-Voigt function over a linear background, until reproducing the linewidth of the reference data and determining the effective polarization of the electron source $P=0.30\pm0.03$.

\subsection{Wavevector resolution}

Once the polarization was obtained from the B1 band, we applied it to the SS at $\theta = 8\degree$ after inelastic background subtraction (Fig. \ref{fig:Fig6}(b)). With calibration of the effective polarization we are able to fairly reproduce the spin asymmetry on data taken with an electron gun of $P = 0.33$ \cite{wissing2013rashba}. The spin asymmetry there from the down-to-up intensity ratio at $k^{\downarrow}_{F}/k^{\uparrow}_{F}$ is $\sim1.6$. Since Zumb\"{u}lte et al. \cite{zumbulte2015momentum} demonstrated that the intensity ratio between $k^{\uparrow}_{F}$ and $k^{\downarrow}_{F}$ is proportional to the electron beam divergence, in our case we have an angular divergence of $\Delta\theta = \pm 3.0\degree$. The angular divergence can be also estimated by considering the spot diameter at the decelerator lens (410 $\mu$m) and at the sample, having $\Delta\theta = \pm 2.3\degree$. We thus obtained a wavevector resolution of $\Delta k_{F}= 0.06$ \AA$^{-1}$ for the setup.

\subsection{\label{sec:level3B}Three-dimensional spin polarization tuning}

In the following, we will show how both the angle with respect to the propagation axis ($\Theta$) and the angle in the plane perpendicular to the propagation ($\Phi$) can be tuned.

\begin{figure}[ht]
	\includegraphics[width=0.5\textwidth]{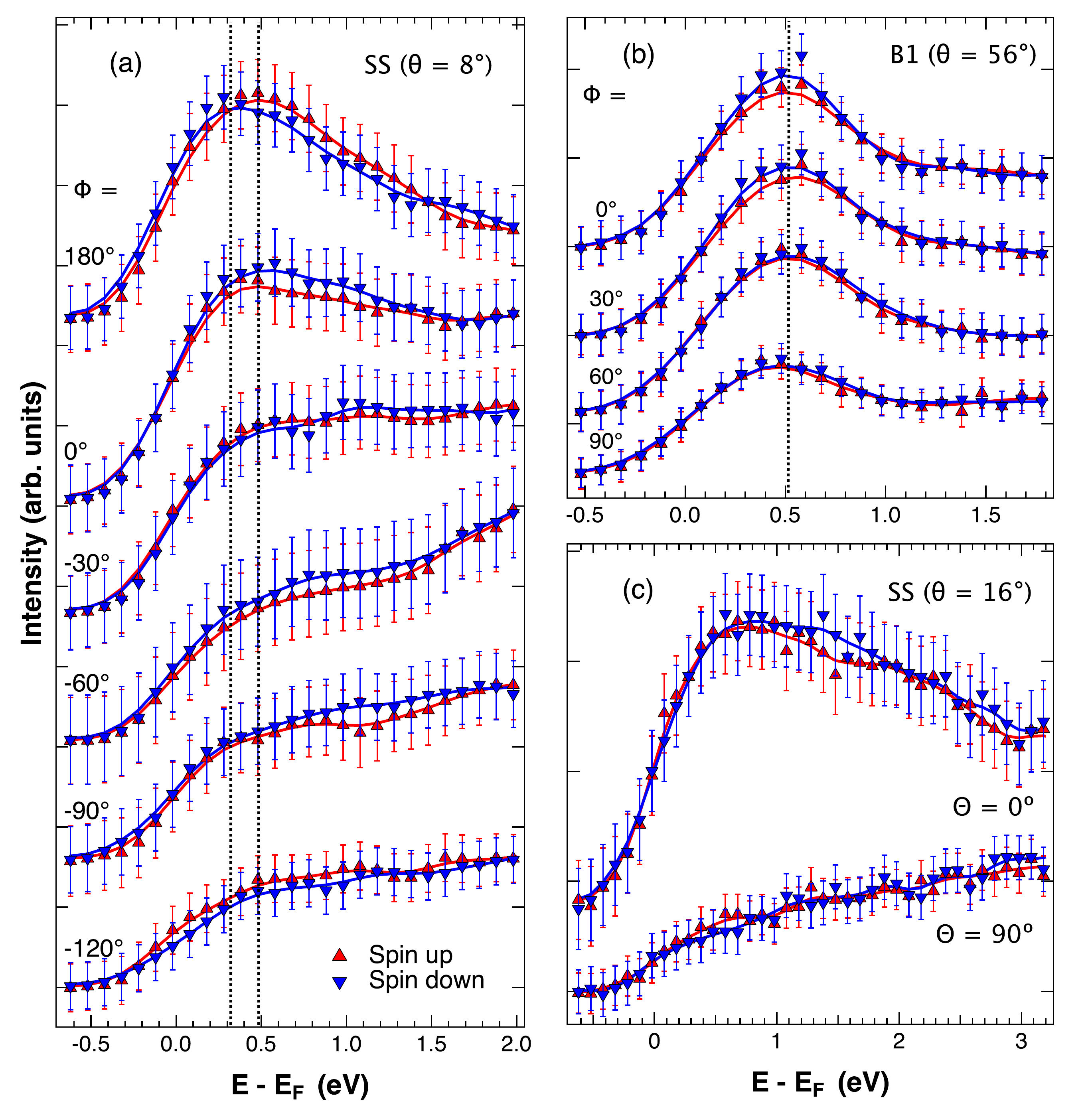} 
	\caption{\label{fig:Fig9} Dependency of Au(111) SPIPES spectra on the transverse ($\Phi$) and longitudinal ($\Theta$) spin polarization with a current density of $\sim$ 0.8 $\mu$A$\cdot$mm$^{-2}$ kept constant by adjusting the power of the laser exciting the photocathode. Smoothed linewidths are shown as guide to the eye. (a) Shockley SS: the inversion between the spin down and spin up components appears after a $\vb{P}$ rotation of $\Phi= 180\degree$ with an energy splitting of $\sim 120$ meV that is within the limits of the experimental uncertainty. (b) The spin asymmetry of B1 vanishes after a rotation of $\Phi = 90\degree$. (c) The Shockley SS intensity decreases for a totally longitudinal component of $\vb{P}$ ($\Theta=90\degree$). Error bars correspond to the standard deviation at each point.}. 
\end{figure}

In order to demonstrate the control of the beam polarization in the plane perpendicular to the beam propagation axis, we studied the Au(111) Shockley SS. This SS has an in-plane spin orientation tangential to the Fermi surface. We have studied the dependence of the experimental spectra when aligning the beam polarization with the SS spin as well as for other relative orientations. Fig. \ref{fig:Fig9}(a) shows the SPIPES spectra of the SS at $\sim k_F$ ($\theta = 8\degree$) for rotations of $\vb{P}$ within the plane perpendicular to the propagation. It can be observed that the spectral intensity as well as the difference between spin up and spin down states increase when the beam polarization is aligned to the SS spin direction. This result is furthermore reproduced for a $180\degree$ rotation and, as expected, the energy splitting $\Delta E = 120$ meV is preserved although the down and up spin peaks exchange their binding energy with respect to $\Phi = 0\degree$. On the other hand, when the beam polarization is perpendicular to the SS spin ($\Phi = -90\degree$), the overall intensity decreases and the energy splitting cannot be observed. In between these extreme situations, the spectral lineshape evolves smoothly. 

\begin{figure}[ht]
	\includegraphics[width=0.5\textwidth]{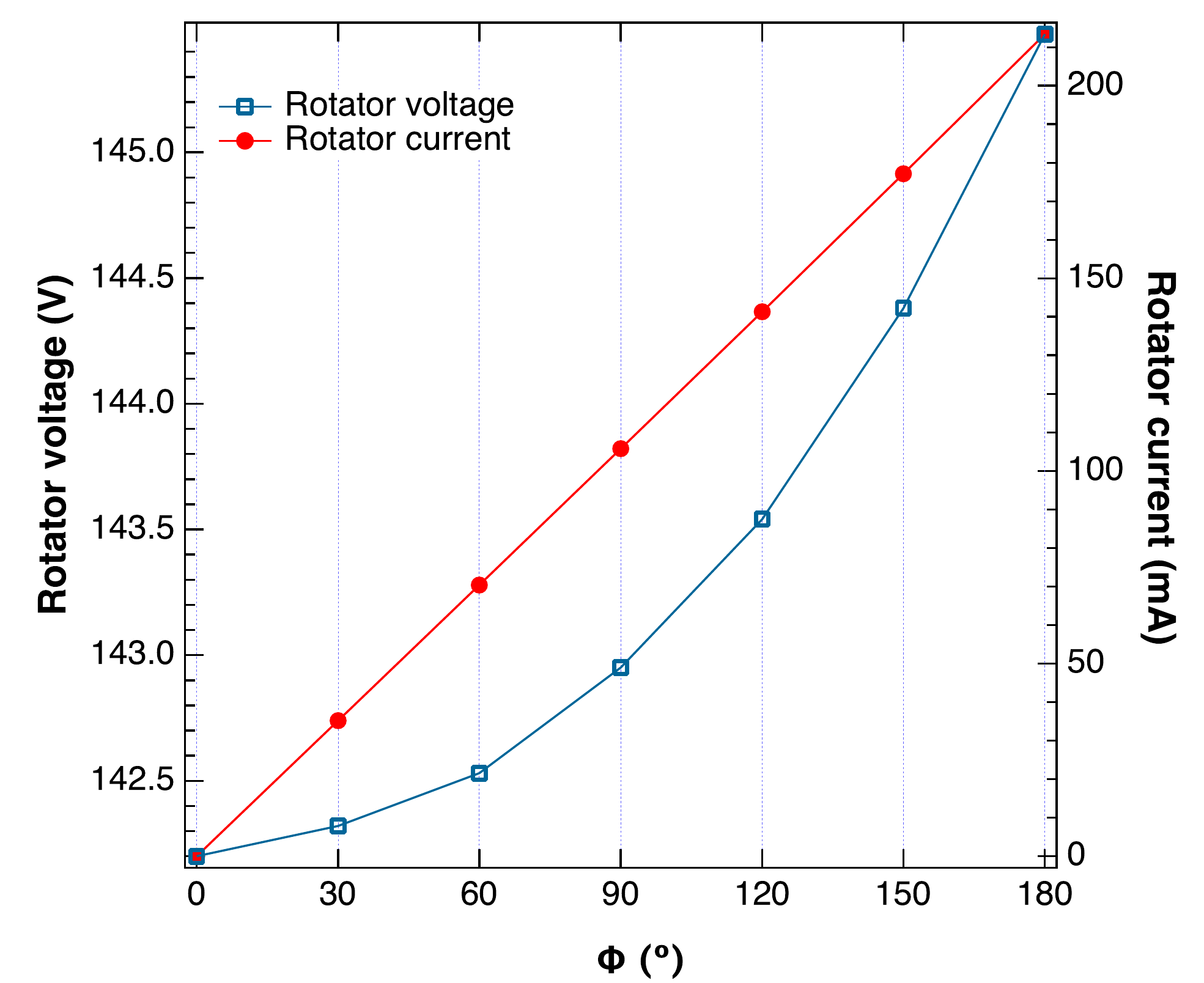} 
	\caption{\label{fig:Fig10} Experimental parameters of the rotator lens for transverse polarization tuning. Both the sign of the rotator current and the sign of $\Phi$ have been set to positive.} 
\end{figure}

Further proof of $\Phi$ tuning is found in Fig. \ref{fig:Fig9}(b). Here we focus on the B1 state of Au(111) at $\theta = 56 \degree$ instead of the Shockley SS. The figure shows how the spin asymmetry is maximized at $\Phi = 0\degree$ and minimized at $\Phi = 90\degree$, as expected. The parameters of the rotator lens that were used for acquiring Fig. \ref{fig:Fig9}(a)-(b) are presented in Fig. \ref{fig:Fig10}.  The rotator current evolves linearly with $\Phi$ whereas the rotator voltage follows a parabolic trend. 

Finally, the in-plane spin of the Au(111) surface state can be used to double check the spin tuning between transverse to longitudinal orientations (Fig. \ref{fig:Fig9}(c)). The SS intensity decreases for longitudinal spin orientation ($\Theta = 90\degree$) but it emerges when retrieving a transverse polarization ($\Theta = 0\degree$), i.e. when the electron beam spin is in-plane, aligned with the SS spin. All these results directly demonstrate the 3D tuning of the spin polarization vector of the electron beam.

Superimposed to an overall beam intensity when tuning the spin, visible in the background, the Au(111) surface state intensity decreases for transverse polarization. This intensity decrease has been observed in photoemission on Au(111) \cite{tusche2015spin}. Quantitative intensity decrease agreement between both techniques is not expected since the experimental geometries in both setups differ and will affect the matrix elements and therefore intensities. Alternatively, spin-torque effect \cite{Slonczewski96, Slonczewski99, Stiles02} also explains the intensity difference for spectra measured with differently oriented incoming spins, due to the different coupling of electrons coming from the vacuum to an electronic band in a ferromagnet. Since the transmitted electron beam into the solid depends on the spin orientation, IPES spectra intensities (but not $E(k)$ relationships) will therefore depend on the spin orientation. A similar intensity decrease of the surface state has also been observed by inverse photoemission when tuning the spin on Tl/Si(111) at $\theta=70\degree$ \cite{Stolwijk2013}. The decrease of intensity when tuning the spin direction is of course not relevant for determining $E(k)$ relationships. Moreover, our setup determines $E(k_{\parallel})$ dispersions for every spin orientation directly, without macroscopically rotating the photocathode and without needing to decompose spectra with coupled in- and out-of-plane spin components through a delicate analysis of spectral intensities. Spectral intensities for the different spin orientations do not need to be compared to obtain meaningful $E(k)$ relationships. 
 
\section{\label{sec:level4}SUMMARY}

A GaAs-based electron gun with a total 3D control of the polarization direction has been adapted with the goal of determining $E(k)$ dispersions in spin-resolved inverse photoemission experiments. The SPIPES source has 30\% polarization and it is able to maintain parallel beam condition and spot location over the target with any desired orientation of the electron beam polarization vector. The wavevector resolution allows to perform angle-resolved inverse photoemission experiments, as shown by measuring the dispersion of Au(111) along $\overline{\Gamma M}$, furthermore demonstrating the capability of the source to measure the Rashba splitting on the Shockley SS. These first SPIPES results demonstrate the performance on this new type of source.  Moreover, we must highlight that fully decoupling the polarization direction from the electron beam direction is a qualitative advance in the field. This decoupling allows to perform new measurements that were not  feasible before, in particular: (1) to perform angle-resolved IPES and explore an arbitrary spin polarization of the initial state at an arbitrary wavevector $k$ (in particular close to $\Gamma$). This is a major advance since existing setups cannot directly explore spins perpendicular to the surface (out-of-plane) except in high values of reciprocal space for arbitrary systems, or when it is expected by symmetry considerations that the in-plane component vanishes\cite{Stolwijk2013,stolwijk2014rotatable}. (2) to measure in- and out-of-plane spectra in fully independent measurements for arbitrary systems, so no additional data treatment involving spectral intensities is necessary to obtain the out-of-plane component. (3) to measure $k$-resolved spectra while keeping constant the polarization projection of the electron beam at every $k$.  We believe that this setup  opens wide perspectives for studying complicated unoccupied-band spin textures.

\section{Supplementary Material}
S1: $E(k)$ dispersion of Au(111) along $\overline{\Gamma}\overline{M}$. S2: Determination of the effective polarization. S3: Comparison of spin-tuning effect on the Au(111) Shockley surface state to literature. S4: Procedure of data  normalization with effective polarization. 

\section*{\label{sec:level5}Data availability statement}
The data described in this article are available from the corresponding author upon reasonable request.

\begin{acknowledgments}
The authors strongly acknowledge CNRS (Institut de Physique)  for funding the spin- and angle-resolved inverse photoemission project, as well as Labex Palm - ANR-10-LABX-0039-PALM - for extra funds (SPOUSE and IP-OP projects). We are grateful to CNRS and Universit\'e Paris-Saclay for providing TD with an invited professor position. We also thank J. de la Figuera, M. Donath, S. Ravy, A. Taleb-Ibrahimi, A.K. Schmid, J. Peretti, Y. Lassailly, E. Ortega, M. A. Gonzalez and J. Caillaux for helpful discussions. We are grateful to A. Bendounan, S. Rohart and A. Thiaville for providing well-characterized test samples and to J. Obando for helping with the optical system. We are extremely grateful for the technical support from M. Bottineau, C. Courtot, V. Davouloury and I. Nimaga (LPS) and also from LAC and ISMO institutes (Orsay, France). 
\end{acknowledgments}

\bibliography{aipsamp}

\end{document}